\documentclass[a4paper,11pt]{article}
\pdfoutput=1 

\usepackage{jinstpub}

\usepackage[utf8]{inputenc}

\usepackage[english]{babel}

\usepackage{float}

\usepackage{multirow, multicol}
\usepackage{makecell}

\usepackage{xcolor} 
\usepackage[normalem]{ulem}

\newcommand{\Xadd}[1]{#1}
\newcommand{\Xreplace}[2]{#2}
\newcommand{\Xremove}[1]{}

\title{\boldmath Study of visible-light emission in pure and methane-doped liquid argon }

\author[a,b]{A.~Bondar,}
\author[a,b]{E.~Borisova,}
\author[a,b]{A.~Buzulutskov,}
\author[a,b]{E.~Frolov,}
\author[a,b]{V.~Nosov,}
\author[a,b,1]{V.~Oleynikov,\note{Corresponding author.}}
\author[a,b]{A.~Sokolov}

\affiliation[a]{Budker Institute of Nuclear Physics, Lavrentiev ave. 11, Novosibirsk 630090, Russia}
\affiliation[b]{Novosibirsk State University, Pirogova st. 2, Novosibirsk 630090, Russia}

\emailAdd{A.F.Buzulutskov@inp.nsk.su}
\emailAdd{V.P.Oleynikov@inp.nsk.su}

\abstract{	
	In liquid argon TPCs for dark matter search and neutrino detection experiments, primary scintillation light is used as a prompt signal of particle scattering, being intensively produced in the vacuum ultraviolet (VUV) due to excimer emission mechanism. On the other hand, there were indications on the production of visible-light emission in liquid argon, albeit at a much lower intensity, the origin of which is still not clear. The closely related issue is visible-light emission in liquid argon doped with methane, the interest in which is due to the possible use in neutron veto detectors for those experiments. In this work we study in detail the properties of such light emission in pure liquid argon and its mixtures with methane. In particular, the absolute photon yield of visible-light emission in pure liquid argon was measured to be about 200 and 90 photon/MeV for X-rays and alpha particles respectively. In liquid argon doped with methane the photon yield dropped down significantly, by about an order of magnitude at a methane molar content varying from 0.01 to 1\%, and then almost did not change when further increasing the methane content up to 10\%.	
}


\begin{document}

	\maketitle
	\flushbottom
	\newcommand*{\doi}[1]{\href{http://dx.doi.org/#1}{doi: #1}}



\section{Introduction}\label{Introduction}

%


In liquid Ar TPCs for dark matter search~\cite{Chepel13} and neutrino detection~\cite{Majumdar21} experiments, the scattered particle produces two types of signals: that of primary scintillation recorded promptly (``S1'') and that of primary ionization recorded with a delay (``S2''). The \Xreplace{ionization (S2)}{S2} signal is recorded typically indirectly in the gas phase of two-phase detectors~\cite{Akimov21} using the effect of proportional electroluminescence~\cite{Buzulutskov20}. In contrast, the \Xreplace{primary scintillation signal (S1)}{S1 signal} is recorded directly using  either photomultiplier tubes (PMTs) or silicon photomultipliers (SiPMs). 

``Ordinary'' primary scintillation light in liquid Ar forms from the excimer (Ar$_{2}^{*}$) emission mechanism~\cite{Chepel13,Akimov21,Buzulutskov17}. Scintillation light is intensely produced in the vacuum ultraviolet (VUV), around 128~nm, with a photon yield of about $40 \times 10^3$~photon/MeV~\cite{Akimov21}, and thus require a wavelength shifter (WLS), such as tetraphenyl-butadiene (TPB), to convert the VUV to the visible light suitable for detection with PMTs and SiPMs.   

On the other hand, there were indications on light emission in the visible and near infrared (NIR) range in liquid Ar in S1 signal~\cite{Buzulutskov11,Bondar12P1,Bondar12P2,Heindl10,Heindl11,Alexander16,Auger16,Escobar18,Bondar20S1}, albeit at a much lower yield, of the order of 500~photon/MeV~\cite{Buzulutskov11}, the origin of which is still not clear. 

It was suggested in ref.~\cite{Buzulutskov18} that such visible-light emission in liquid Ar might be explained by neutral bremsstrahlung (NBrS) of primary ionization electrons, decelerated in the medium down to the energy domain of the NBrS effect (1-10~eV). 
The NBrS effect is that of bremsstrahlung of electrons scattered on neutral atoms \cite{Buzulutskov18}:
\begin{eqnarray}
	\label{Rea-NBrS-el}
	e^- + \mathrm{A} \rightarrow e^- + \mathrm{A} + h\nu \; . 
\end{eqnarray}
So far, the NBrS emission in the form of electroluminescence in noble gases has been experimentally and theoretically studied in refs.~\cite{Buzulutskov18,Bondar20NBrS,Kimura20,Takeda20,Aoyama21,Aalseth21,Henriques22} and~\cite{Buzulutskov18,Borisova21G,Borisova21L,Amedo22} respectively, with predicted continuous spectrum in the visible and NIR range.

\Xadd{The NBrS emission in single collision is naturally fast as it is defined by time of flight of the electron (of 1-10~eV kinetic energy) nearby the atom, which is of the order of $10^{-4}$~ps. Since to reduce the electron energy from 10 to 1~eV it takes about $10^5$ collisions, with electron mean free path between them of about 10~nm and electron velocity of about $10^{6}$~m/s, the overall time of the NBrS  emission from primary ionization electrons would be of the order of 1~ns.} 

\Xadd{Also} it is known that electroluminescence in the VUV in gaseous Ar due to the excimer effect is associated with primary scintillation VUV light in liquid Ar due to the same excimer effect. \Xadd{Similarly, one could assume that the visible-light emission due to the NBrS effect would exist both in gaseous and liquid Ar.} \Xremove{Accordingly, why should this not happen with NBrS emission? Namely, since electroluminescence in gaseous Ar in the visible range do exist due to the NBrS effect, their analog in liquid Ar should definitely exist, again in the visible range, in the form of visible-light emission of S1 signal due to the same NBrS effect.} 

The closely related issue is visible-light emission in liquid Ar doped with CH$_4$, the interest in which is due to the possible use in neutron veto detectors for dark matter search experiments.  To achieve a high efficiency of neutron detection, the working medium of the veto detector should contain hydrogen, acting as a neutron moderator. So far, neutron veto detectors based on flammable mixtures of liquid organic scintillators have been used~\cite{Agnes15,Agnes16}.  
A safe alternative might be a liquid scintillator based on liquid Ar doped with CH$_4$. This idea was discussed for application in the DarkSide experiment~\cite{Galbiati_Private_communication}. In particular, the CH$_4$ content of 5-10\% in liquid Ar would be enough for compact neutron veto detector of about 1~m thick.

However, it is well known that even tiny amount ($>$0.1~ppm) of CH$_4$ dopant would immediately quench excimer scintillations in liquid Ar~\cite{Jones13}. The quenching is due to both VUV light absorption on CH$_4$~\cite{Jones13,SpectralAtlas} and excimer deexcitation in collisions with CH$_4$ molecules. On the other hand, this may not be the case for visible-light emission due to the NBrS effect: it can be expected not to be suppressed by the CH$_4$ dopant.  Indeed, the NBrS radiation induced by primary ionization electrons was supposed to be present not only in atomic noble gases, but in molecular gases as well, namely in air~\cite{AlSamarai16}. Similarly,
we may suppose that the NBrS emission could also exist in a liquid mixture of Ar and CH$_4$ in the form of light emission in the visible and NIR range.

Accordingly, in this work we systematically study the properties of visible-light emission of S1 signal both in pure liquid Ar and its mixtures with CH$_4$ to verify the previous results and hypotheses. The preliminary results of the study were published in ref.~\cite{Bondar20S1}.
In the current work, the experimental setup design was improved compared to that of ref.~\cite{Bondar20S1}, namely photon collection efficiency was increased, cross-talk background was decreased and an alpha-particle source of irradiation was used  in addition to that of pulsed X-ray.
As a result, the photon yields of visible-light emission were measured in pure liquid Ar and its mixtures with CH$_4$ with improved accuracy. The relevance of the results obtained to the development of noble liquid detectors for dark matter searches is also discussed.

\section{Experimental setup}

\begin{figure}[!htb]
	\center{\includegraphics[width=0.7\columnwidth]{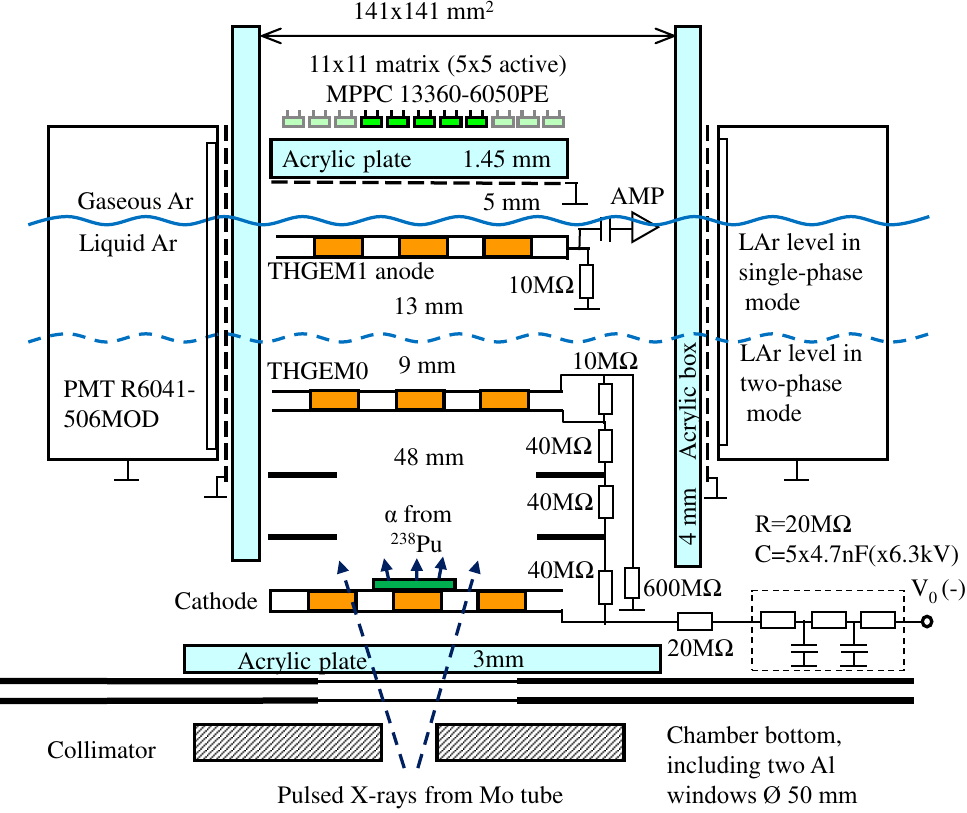}}
	\caption{Schematic view of the experimental setup.}
	\label{fig_setup}
\end{figure}

Fig.~\ref{fig_setup} shows the experimental setup. The cryogenic chamber was operated in a single-phase or two-phase mode: it was filled with 3.5 or 2.5~liters of the liquid (pure Ar or Ar doped with CH$_4$) respectively.
The detector was operated in equilibrium state at a saturated vapor pressure of 1.00~atm. For pure Ar this corresponds to a temperature of 87.3~K. For Ar+CH$_4$ mixture, the actual temperature depends on the CH$_4$ content.

The first measurement sessions were done with pure Ar.
To prepare a certain Ar+CH$_4$ mixture before the next sessions, the gas composition was changed by cooling the bottle with given mixture and adding the required amount of CH$_4$ or pure Ar.
The CH$_4$ content in the mixture prepared this way was measured using a Residual Gas Analyzer (RGA) Pfeiffer-Vacuum QME220 F2~\cite{RGA} in a flow mode at a pressure reaching $10^{-4}$~mbar.
The difference between the expected and measured content values was below 10\%.

The measurements were performed using Ar gas of an initial purity of 99.9998\% (with the maximum impurity levels [N$_2$]<0.8~ppm, [O$_2$]<0.5~ppm, [H$_2$O]<0.5~ppm) and CH$_4$ of an initial purity of 99.95\% (with the maximum impurity levels [N$_2$]<200~ppm, [O$_2$]<10~ppm, [H$_2$O]<10~ppm). At the beginning of each measurement session, the chamber is evacuated and filled with pure argon. After repeating this procedure several times, the content of residual impurities in the chamber decreases and we begin gas liquefaction. The gas mixture from the stainless steel bottle was passed through an Oxisorb filter~\cite{spectron} for purification from electronegative impurities and then was liquefied into the cryogenic chamber. By purifying the gas mixture with the Oxisorb filter, the content of O$_2$ and H$_2$O was reduced down to a few ppb, according to charge measurements (see section~\ref{Energy_deposition}). At the end of the session, the content of the cryogenic chamber was collected back into the bottle.

In measurements with pure Ar the N$_2$ content was below 1~ppm (with the measurement accuracy of 1~ppm); it was monitored before and after  measurement session using an ``SVET'' gas analyzer~\cite{svet}, which employed emission-spectrum-measurement technique.

The cryogenic chamber operated in the single-phase mode was composed of two gaps: that of drift (with low electric field, namely drift field $F_d$), 48~mm thick, and that of induction (with high electric field, $F_i$), 22~mm thick. 
To form these gaps, the electrodes made from THGEMs (Thick Gas Electron Multipliers,~\cite{Breskin09}) were used: see Fig.~\ref{fig_setup}. 
The drift gap was formed by a cathode electrode, field-shaping electrodes and THGEM0. 
The induction gap was formed by THGEM0 and THGEM1, the latter acting as an anode.
In the two-phase mode the induction gap was divided into two: that of extraction (with high electric field, $F_{extr}$), 9~mm thick, located between THGEM0 and liquid level, and that of electroluminescence (with high electric field, $F_{EL}$), 13~mm thick, located between the liquid level and THGEM1.
The liquid level was calculated from the amount of condensed Ar using CAD software and was verified using THGEM1 as a capacitive liquid level meter.

Two different types of spectral devices were used in the measurements.
Four compact cryogenic PMTs R6041-506MOD~\cite{Bondar15PMT} were located on the perimeter of the induction (or extraction/electroluminescence) gap and electrically insulated from it by an acrylic box.
Another spectral device was a 5$\times$5 SiPM matrix, composed from SiPMs of MPPCs 13360-6050PE type~\cite{hamamatsu} with an active area of 6$\times$6~mm$^2$ each and channel pitch of 1~cm.
The PMTs and SiPM matrix were protected from the TPC electric fields using grounded metal meshes in front of them. The independence of their gain characteristics from the electric field was confirmed in experiment.

\begin{figure}[!htp]
	\center{\includegraphics[width=0.7\columnwidth]{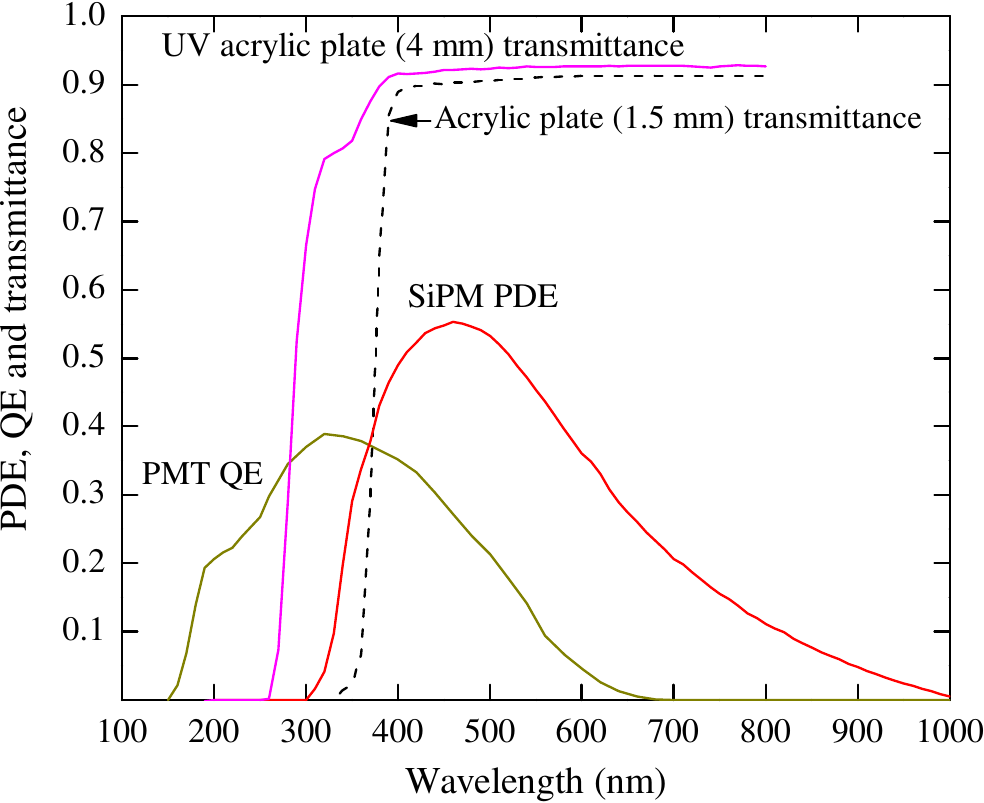} \\ \vspace{5mm}
	\includegraphics[width=0.7\columnwidth]{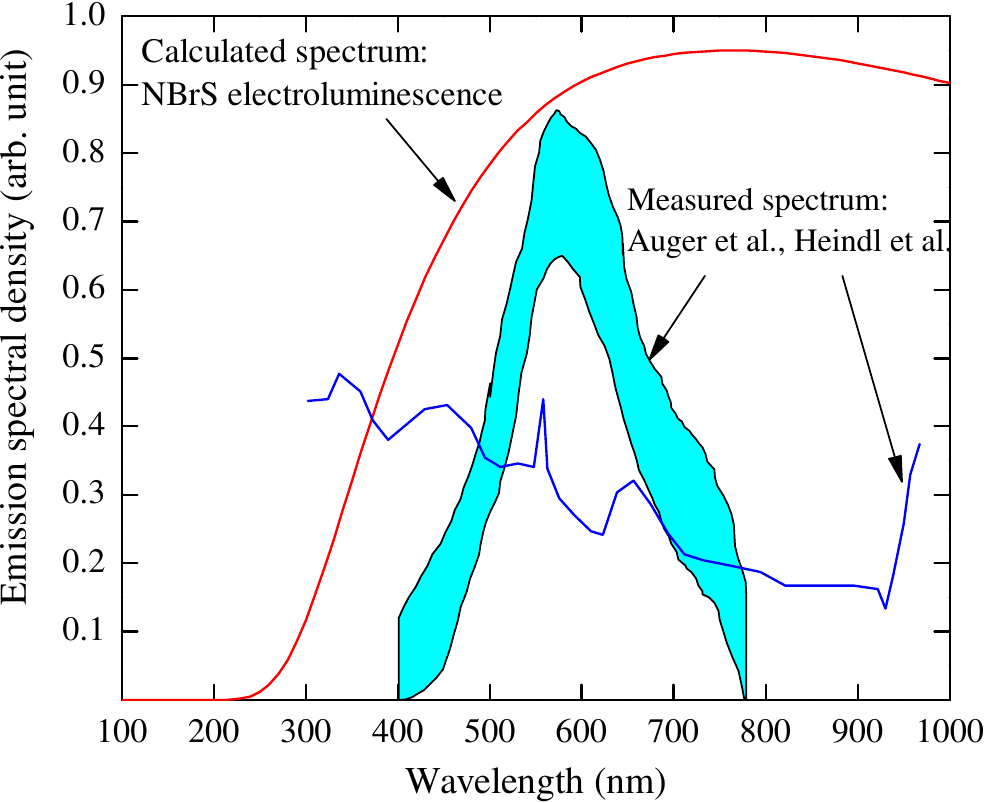}}
	\caption{Top: Photon Detection Efficiency (PDE) of SiPM (MPPC 13360-6050PE~\cite{hamamatsu}) at overvoltage of 5.6~V obtained from ref.~\cite{Otte17} using the PDE voltage dependence, Quantum Efficiency (QE) of the PMT R6041-506MOD at 87~K obtained from refs.~\cite{hamamatsu,Lyashenko14} using a temperature dependence derived there, the transmittance of the UV and ordinary acrylic plate in front of the PMTs and SiPM matrix, respectively. Bottom: liquid Ar emission spectra measured in Heindl et al.~\cite{Heindl11} (light emission of S1 signal) and Auger et al.~\cite{Auger16} (electroluminescence due to HV breakdown at $\leq 200$~kV/cm), as well as electroluminescence spectrum due to the NBrS EL effect at 200 kV/cm theoretically calculated in ref.~\cite{Borisova21L}.}
	\label{fig_spectra}
\end{figure}

Fig.~\ref{fig_spectra} (top) shows the spectral characteristics of the detector, namely the Photon Detection Efficiency (PDE) of SiPMs, Quantum Efficiency (QE) of PMTs, transmittance of the UV and ordinary acrylic plate in front of the PMTs and SiPM matrix, respectively.
Taking into account the transmission of the acrylic plates, the PMTs and SiPM matrix were sensitive in the range of 270-650~nm and 360-1000~nm, respectively.
The spectral devices were insensitive to the VUV, since no WLS was used in the experimental setup.

For comparison, Fig.~\ref{fig_spectra} (bottom) shows the emission spectra in the visible range in pure liquid Ar, measured in Heindl et al.~\cite{Heindl11} and Auger et al.~\cite{Auger16} for light emission of S1 signal and electroluminescence respectively. We also show the spectrum of electroluminescence in liquid Ar due to the NBrS EL effect, theoretically calculated in ref.~\cite{Borisova21L} for the electric field of 200 kV/cm, the latter roughly corresponding to the maximum field used in Auger et al.~\cite{Auger16}. One can see that all the spectra are rather flat. Also, both the SiPM matrix and PMTs were sensitive to all of them.

The detector was irradiated from outside by X-rays from a pulsed X-ray tube with Mo anode, with the average deposited energy in liquid Ar of 25~keV~\cite{Bondar16}, or by alpha particles with the energy of 5.5~MeV from a $^{238}$Pu source with an activity of about $5\times10^4$~Bq, installed at the center of the cathode inside the detector.

The signals from the PMTs were amplified using fast 10-fold amplifiers CAEN N979 and linear amplifiers with a shaping time of \Xreplace{200~ns}{10 and 200~ns respectively}.
The signals from SiPMs were transmitted to fast amplifiers with a shaping time of 40~ns, via twisted pair wires. The charge signal from the THGEM1 anode was recorded using a calibrated chain of a preamplifier and shaping amplifier. All amplifiers were placed outside the detector.

\begin{figure}[!htp]
	\center{\includegraphics[width=0.70\columnwidth]{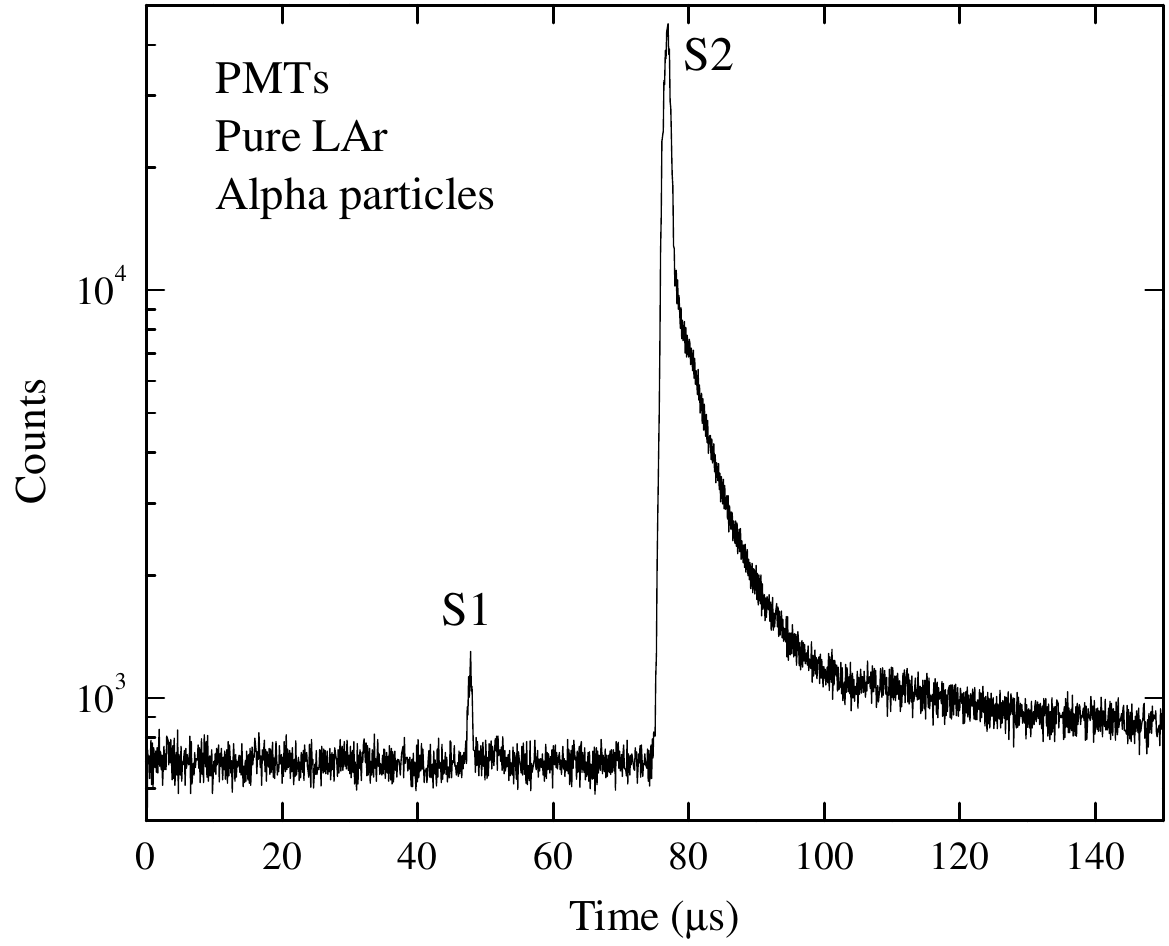}}
	\caption{Pulse-shape of S1 and S2 signals, induced by alpha particles, obtained in the two-phase mode and recorded by PMTs.
		The data were obtained at the maximum electric field in pure liquid Ar ($F_d=0.62$~kV/cm, $F_{EL}=7.8$~kV/cm).
		The S2 signal amplitude significantly exceeded that of S1,  which allowed to use the S2 signal as a trigger. Note that the S2 signal has the slow component with time constant of about 5~$\mu$s, similar to that observed elsewhere \cite{Bondar20S2}.}	
	\label{image:fig_time_spectrum_alpha_noTPB}
\end{figure}

The DAQ system included both a 4-channel oscilloscope LeCroy WR HRO 66Zi and a 64-channel Flash ADC CAEN V1740 (12~bits, 62.5~MHz): the signals were digitized and stored both in the oscilloscope and in a computer for further off-line analysis.
In measurements with the pulsed X-ray tube, the trigger was provided by its pulse generator. In measurements with alpha particles, the trigger was provided by the S2 signal taken from all the PMTs, in this case the detector being operated in the two-phase mode: see Fig.~\ref{image:fig_time_spectrum_alpha_noTPB}. Note that at higher electric fields the electroluminescence (S2) signal had the characteristic slow component with time constant of about 5~$\mu$s, similar to that observed elsewhere \cite{Bondar20S2}.

\section{Measurements with pure liquid argon}\label{section_pure_liquid_argon}
\subsection{Pulse-shape analysis}

\begin{figure}[h!]
	\center{\includegraphics[width=0.85\columnwidth]{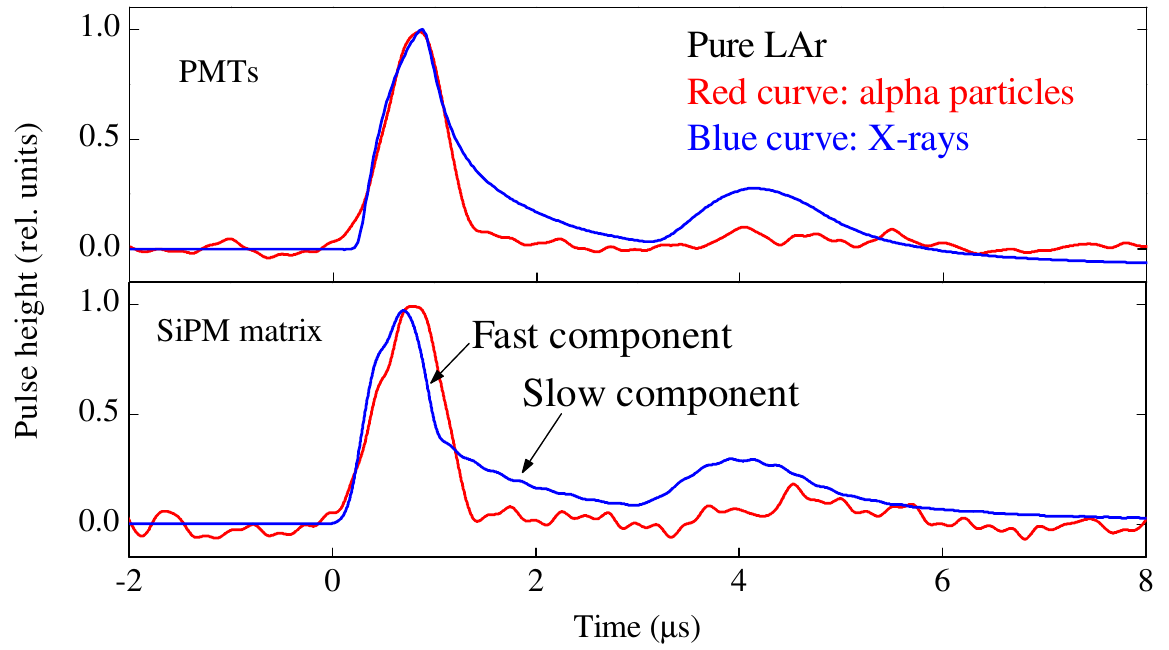}}
	\caption{Averaged pulse shapes of the S1 signals, induced by alpha particles and pulsed X-rays, obtained in the two-phase mode in pure liquid Ar and recorded by SiPM matrix and PMTs. Data with pulsed X-rays and alpha particles were obtained at zero and maximum electric field (F$_d= 0.62$~kV/cm, F$_{EL} = 7.8$~kV/cm), respectively. All the pulse heights are normalized to the maximum.}
	\label{image:fig_pulse_shape_alpha_vs_xray_noTPB}
\end{figure}

Pulse shape analysis can provide valuable information on the mechanisms of light emission in liquid Ar. Fig.~\ref{image:fig_pulse_shape_alpha_vs_xray_noTPB} shows the averaged pulse shape of the S1 signal from PMTs and SiPM matrix, induced by alpha particles and pulsed X-rays, obtained in pure liquid Ar. 

The S1 signal obtained with the pulsed X-ray tube had two distant peaks due to characteristic double-pulse structure of the X-ray tube itself, which was determined in special measurements using a BGO scintillation counter~\cite{Bondar12P1}. The peak width is defined by the X-ray pulse width, of about 0.5~$\mu$s \cite{Bondar12P1}, from which one can estimate the time constants ($\tau$) of visible-light emission themselves. In particular, one can deduce that visible-light emission in pure liquid Ar had the fast ($\tau_{f}<100$~ns) and slow ($\tau_{s}\sim 1~\mu$s) components of about the same contribution. 
On the other hand, the S1 signal obtained with the alpha particle source had the fast component only. 
Also, the S1 pulse shape did not depend on the electric field.

\subsection{Photoelectron yield dependence on the drift field}

\begin{figure}[h]
	\center{\includegraphics[width=0.7\columnwidth]{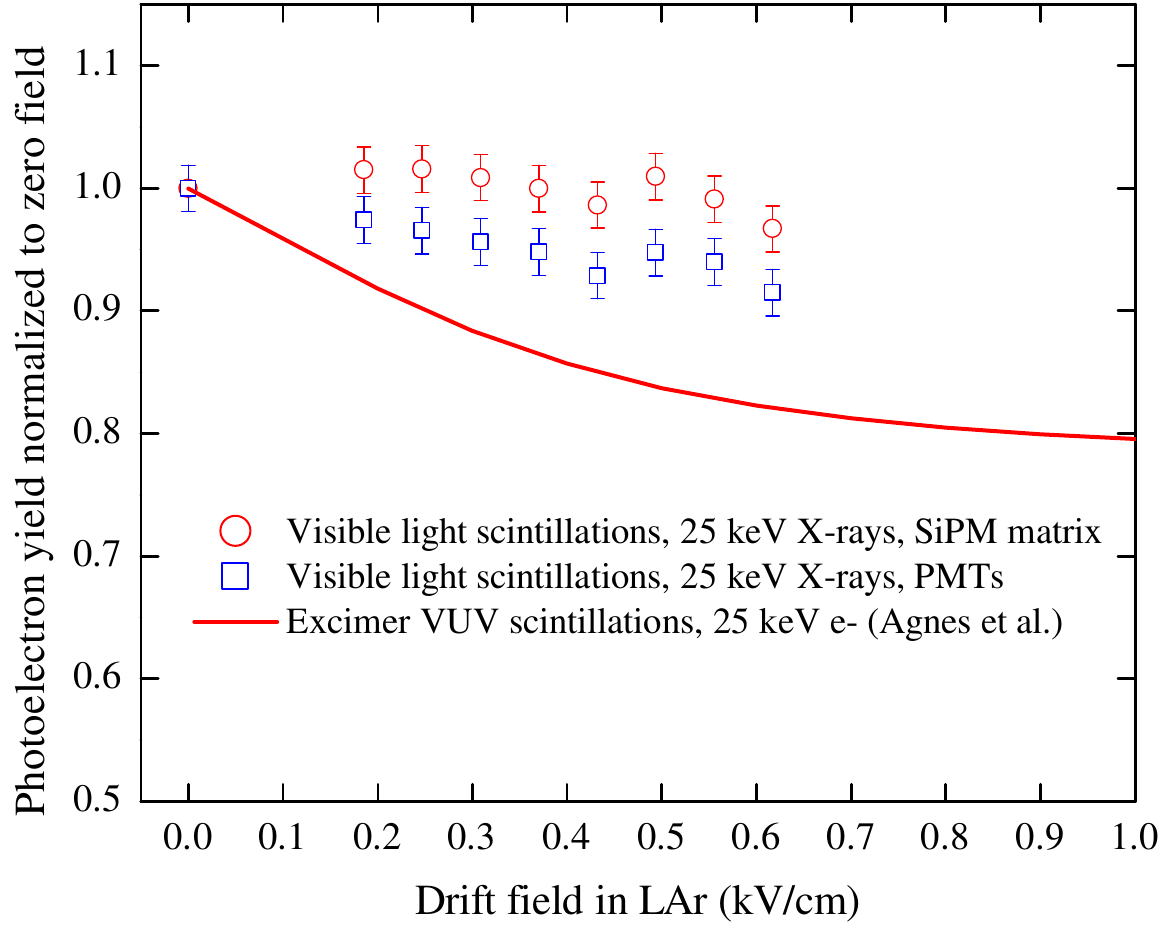}}
	\caption{Relative photoelectron yield of visible-light emission of S1 signal in pure liquid Ar recorded by PMTs and SiPM matrix in single-phase mode, induced by pulsed X-rays with an average deposited energy of 25~keV, as a function of the electric field (data points). For comparison, the relative photoelectron yield of excimer (VUV) scintillations induced by electrons with an energy of 25~keV is shown, obtained from Agnes et al.~\cite{Agnes18NR} using the stopping power of electrons in Ar from ref.~\cite{Berger17} (curve).}
	\label{image:fig_S1_vs_field_xray_noTPB} 
\end{figure}

Fig.~\ref{image:fig_S1_vs_field_xray_noTPB} shows the photoelectron yield of visible-light emission in liquid Ar, expressed in the number of photoelectrons (photoelectron number) recorded by PMTs and SiPM matrix,  as a function of the drift field. The photoelectron number was integrated over the time interval of 10~$\mu$s from the beginning of the signal and normalized to that of zero electric field. One can see that the photoelectron yield of visible-light emission is almost independent of the drift field, in contrast to that of excimer (VUV) scintillations, the yield of which substantially decreases with the drift field, the latter being explained by weakening of the electron-ion recombination at higher electric fields \cite{Akimov21}. 

This difference obviously indicates on the different mechanisms of visible-light emission and excimer scintillation in liquid Ar, in particular on that visible-light emission is not related to excited Ar states (the latter are copiously produced in electron-ion recombination). On the other hand, the light yield independence of the electric field is naturally explained in the frame of NBrS mechanism,  where the photons are emitted due to elastic electron scattering on atoms (see Eq.~\ref{Rea-NBrS-el}). 

It should be noted that a similar plot for alpha particles does not make much sense, because of the small statistics the photoelectrons number fluctuations are too large, of about 10\%, i.e of the order of the effect.

\subsection{Energy deposition}\label{Energy_deposition}

To calculate the absolute photon yield of visible-light emission one need to know the energy deposited in liquid Ar.
In measurements with alpha particles, one alpha particle was mainly recorded per event, resulting in that the deposited energy (5.5~MeV) was well defined.
In measurements with pulsed X-rays, the detector recorded a large number of X-ray photons in each pulse and the energy deposition ($E_{dep}$) can be determined using the following formula:
\begin{equation}
\label{eq_E_dep} E_{dep} = N_{i} \cdot W \, ,
\end{equation}
where $N_{i}$ is the primary ionization charge and $W$=23.6~eV is the energy needed to produce one ion pair in liquid Ar~\cite{Miyajima74}.

The primary ionization charge ($N_i$) was calculated from the dependence of the collected charge ($N_{coll}$) on the electric field ($F$) in liquid Ar, accounting for the recombination effect using the following parameterization with the recombination coefficient $k_{rec}$~\cite{Chepel13,Aprile06,Akimov21}:
\begin{equation}
\label{eq_charge_vs_field_full} N_{coll} = N_{i} \cdot \frac{T_e \cdot \exp(- K_{att} \cdot C \cdot X  )}{1 + k_{rec}/F} \, .
\end{equation}
Here in addition, both the electron transmission through the THGEM0 electrode \textcolor{black}{($T_e=99\%$)} and the attachment of electrons drifting over the distance $X$ to electronegative impurities (with concentration $C$) were taken into account,  using the attachment coefficient $K_{att}$. The latter was taken as~\cite{Aprile06}: 
\begin{equation}
\label{eq_Katt} K_{att} = 0.95 / F^{0.8} \, ,
\end{equation}
where $K_{att}$ is expressed in (ppm$\cdot$mm)$^{-1}$ and $F$ in kV/cm.
The recombination coefficient is a function of the energy and is described by the following equation (see Fig.~4 in~\cite{Bondar20S1}):
\begin{equation}
\label{eq_Krec} k_{rec}\text{[V/cm]} = 485 + 47000 / \text{E[keV]} \, .
\end{equation}

\begin{figure}[h!]
	\center{\includegraphics[width=0.7\columnwidth]{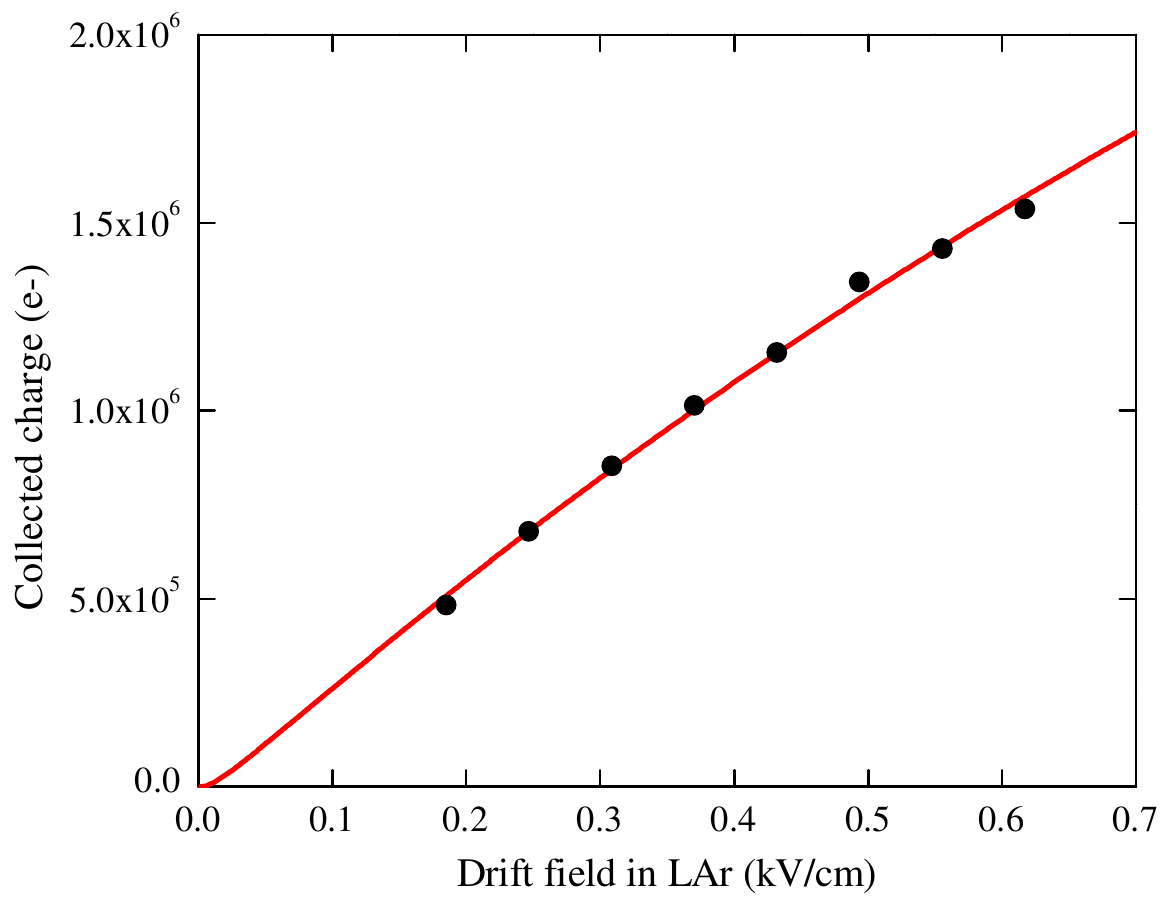}}
	\caption{Collected charge in pure liquid Ar as a function of the drift field, obtained in the single-phase mode. Red curve is the fit by~\eqref{eq_charge_vs_field_full}, where $N_{i}$ and $C$ are free parameters.}
	\label{fig_charge_vs_field_pureAr_210422}
\end{figure}

The collected charge in pure Ar was fitted by~\eqref{eq_charge_vs_field_full}, where the primary ionization charge ($N_i$) and impurity concentration ($C$) were used as free parameters: see Fig.~\ref{fig_charge_vs_field_pureAr_210422}.
The obtained impurity concentration amounted to about $2 \pm 0.8$~ppb, which corresponds to electron lifetime of $150 \pm 60 \ \mu$s at a drift field of 200~V/cm.

The obtained primary ionization charge in measurements with pulsed X-rays amounted to $7.9 \times 10^6$~e$^{-}$, which corresponds to the energy deposition of 186~MeV.

\subsection{Absolute \Xadd{visible} light yield}\label{subsec_LAr_Calc_of_abs_yield}

The absolute light yield \Xremove{(or else photon yield)} is defined as the number of emitted photons per 1 MeV of deposited energy, i.e. it is equal to the ratio of the photoelectron number to the deposited energy and to the photon-to-photoelectron conversion efficiency ($PCE$): $Y = N_{PE} / E_{dep} / PCE$.
The conversion efficiency is defined as $PCE = \varepsilon < PDE >$.
Here $\varepsilon$ is the photon collection efficiency, calculated using Monte Carlo simulation, and $< PDE >$ is the SiPM PDE or PMT QE averaged over the emission spectrum of liquid Ar taken from either \cite{Heindl11} or \cite{Auger16} and appropriately convoluted with the acrylic transmittance spectrum (see Fig.~\ref{fig_spectra}).

\begin{table}[h]
	\caption{Absolute light yield induced by pulsed X-rays and alpha particles, recorded by SiPM matrix and PMTs and calculated in the wavelength range of 400-1000~nm using emission spectrum from either \cite{Heindl11} or \cite{Auger16}.} \label{table_pure_LAr_abs_light_yield}
	\begin{center}
		\begin{tabular}{|cccc|}
			\hline
			\multirow{2}{*}{Source} & \multirow{2}{*}{Spectral device} &  \multicolumn{2}{c|}{Light yield (photon/MeV) for spectrum of } \\
			&  & \cite{Heindl11}  &  \cite{Auger16} \\
			\hline
			\multirow{2}{*}{\makecell{25~keV X-rays}} & PMTs & $71 \pm 14$ & $121 \pm 24$ \\
			  & SiPM matrix &  $212 \pm 42$  &  $187 \pm 37$ \\ \rule{0pt}{4ex}
			\multirow{2}{*}{5.5~MeV $\alpha$} & PMTs &  $43 \pm 9$ &  $107 \pm 22$ \\
			 & SiPM matrix & $98 \pm 20$ & $87 \pm 17$ \\
			\hline
		\end{tabular}
	\end{center}
\end{table}

The light yield thus obtained is shown in table~\ref{table_pure_LAr_abs_light_yield} for pure liquid Ar at the maximum electric field and wavelength range of 400-1000~nm. 
The error in the recorded photoelectron number was about 10\% for X-rays (mainly due to calibration uncertainty) and about 15\% for alpha-particle source (due to calibration uncertainty and low photoelectron statictics). The systematic error in deposited energy was about 15\% for X-rays, while for alpha particles it was insignificant. The PCE error was about 10\% and is mainly due to uncertainty in QE and PDE of PMTs and SiPMs, respectively. Using the root mean square formula to sum the mentioned factors, the error in the absolute light yields turned out to be about 20\%, regardless of the photon detector and radiation source.
Interestingly, the yield is almost independent of which emission spectrum is used, in particular for SiPM matrix data, which indicates that the results are weakly dependent on the shape of the spectrum if the latter is sufficiently flat (as in our case).
Therefore in the following, we use the light yields obtained with the SiPM matrix rather than those of PMTs.  
The resulting light yield in pure liquid Ar was taken as the average between the yields for the two emission spectra (for that of \cite{Heindl11} and \cite{Auger16}): it amounted to $200 \pm 50$~photon/MeV and $92 \pm 23$~photon/MeV for pulsed X-rays and alpha particles, respectively.

\subsection{Comparison with the previous results}

Table~\ref{table_pure_LAr_overview} shows a compilation of the results on light emission in liquid Ar in the visible and NIR range. Let us compare these to those of the present work.

\begin{table}[h]
	\caption{Compilation of the results on light emission in liquid Ar in the visible and NIR range. 
	}
	\label{table_pure_LAr_overview}
	\begin{center}
		\begin{tabular}{|cccccc|}
			\hline
			Reference & \makecell{Excitation \\ source} & \makecell{Electric \\ field \\ (kV/cm)} & \makecell{Spectrum \\ range \\ (nm)} & \makecell{Light yield \\ in visible and \\ NIR range  \\ (photon/MeV)} & \makecell{Comments \\ } \\
			\hline
			\cite{Heindl10,Heindl11} & \makecell{12~keV e$^-$} & 0 & 300-1000 &  Observed & \makecell{Spectrum \\ measured} \\ \rule{0pt}{5ex} 
			\makecell{\cite{Buzulutskov11,Bondar12P1,Bondar12P2}} & \makecell{25~keV \\ X-rays} & 0-30 & 400-1000 & \makecell{$510 \pm 90$ \\ } & \makecell{Field independent} \\ \rule{0pt}{5ex}
			\makecell{\cite{Neumeier14} \\ \cite{Neumeier15}} & 12~keV e$^-$ & 0 & 500-1000 & \makecell{Not observed \\ Observed in NIR} &  \\ \rule{0pt}{5ex}
			\makecell{\cite{Alexander16}} &  \makecell{511~keV \\ $\gamma$-rays} & 0 & 715-900 &  Observed & \makecell{$\tau_{f}$<100~ns \\ $\tau_{s} \approx 2-4$~$\mu$s \\ Field independent}  \\ \rule{0pt}{5ex}
			\makecell{\cite{Auger16}} & \makecell{HV breakdown \\ } & $\leq$200 & 400-800 &  Observed & \makecell{Spectrum \\ measured} \\ \rule{0pt}{5ex}
			This work & \makecell{25~keV \\ X-rays} & 0-0.62 & 400-1000 &\makecell{ $200 \pm 50$ \\ }  & \makecell{$\tau_{f}$<100~ns \\  $\tau_{s} = 1 \pm 0.3$~$\mu$s} \\ 			
			\hline
			\makecell{\cite{Jones13}}& 5.3~MeV $\alpha$ & 0 & 300-650 & \makecell{Not \\ observed} & \makecell{<10 ph./MeV}  \\ \rule{0pt}{5ex}
			\makecell{\cite{Hofmann13}} & 10 MeV protons & 0 & 300-1000 & Observed &  \makecell{Spectrum \\ measured} \\ 
			\rule{0pt}{5ex}											
			\makecell{\cite{Escobar18}} & 5.4~MeV $\alpha$ & 0 & 715-900 & Observed &  \\ 
			\rule{0pt}{5ex}
			This work & 5.5~MeV $\alpha$ & 0.3-0.62 & 400-1000 & \makecell{$92 \pm 23$ } & \makecell{$\tau_{f}$<100~ns \\	
				No slow comp. \\ Field independent} \\		
			\hline
		\end{tabular}
	\end{center}
\end{table}

First of all, the results of the present work confirm those of our (Novosibirsk group) previous works \cite{Buzulutskov11,Bondar12P1,Bondar12P2}. Indeed, visible-light emission in liquid Ar was observed there in the range of 400-1000~nm, using the same X-ray tube. The signal pulse shapes were analyzed on a large time scale, of about 50~$\mu$s, resulting in that the fast and slow components were not distinguishable, unlike the measurements of the present work (see Fig.~\ref{image:fig_pulse_shape_alpha_vs_xray_noTPB}). 
Using the spectrum from~\cite{Heindl11}, the absolute light yield was determined: it amounted to $510 \pm 90$~photon/MeV, which is consistent with that of the present work within a factor of 2.5. Similarly to the present work, almost no field dependence was observed.

It should be remarked that in the present work, in measurements with X-rays the signal had the fast ($\tau_{f}<100$~ns) and slow ($\tau_{s}\sim 1~\mu$s) components with approximately the same contribution. On the other hand, 
in measurements with alpha particles, only the fast component was observed, which may indicate on somewhat different emission mechanisms when irradiated by X-rays and alpha particles.

The present results are also in compliance with those of~\cite{Alexander16,Escobar18} (FNAL group): light emission in the NIR in liquid Ar were observed there in the range of 715-900~nm, when irradiated with 511~keV gamma-rays and 5.4~MeV alpha particles. 
Analyzing the pulse shapes presented in~\cite{Alexander16}, one may conclude that the authors observed the fast ($\tau_{f}<100$~ns) and slow ($\tau_{s}=2-4~\mu$s) signal components, similarly to that of the present work. 

The third group that observed visible-light emission in liquid Ar was that of \cite{Auger16}: electroluminescence was observed at 400-800~nm, induced by high-voltage electrical breakdown. The spectrum of the emission continuum was measured.

The results of the fourth group (Munich group) are somewhat contradictory. On one hand, an  emission continuum in the visible and NIR range was observed in liquid Ar at 300-900~nm \cite{Heindl10,Heindl11,Hofmann13}, in addition to excimer scintillations in the VUV, when irradiated with 12 keV electron and 10 MeV proton beams.  The spectrum of the continuum was measured. Moreover, in \cite{Hofmann13} it was hypothesized that this emission continuum could be explained by the bremsstrahlung effect. 
On the other hand, later the same group \cite{Neumeier14,Neumeier15} first claimed that the continuum observed in~\cite{Heindl10,Heindl11,Hofmann13} ``was an artifact due to the normalization of the spectrum with the response function of the spectrometer used for that spectral region'' \cite{Neumeier14}, but then again reported on observation of emission in the NIR \cite{Neumeier15}. Therefore, we tend to think that this group did observe such visible-light emission and therefore it is justified to use their spectrum in the analysis of the results.

Finally, the fifth group \cite{Jones13} did not observe photon emission in liquid Ar in the visible range, when irradiated with 5.3~MeV alpha particles, in contradiction with the results of our and FNAL groups: the light yield was estimated to be less than 10~photon/MeV, which is an order of magnitude lower than that of the present work.

\section{Measurements with liquid argon doped with methane}\label{section_argon_doped_with_methane}
\subsection{Pulse-shape analysis}

\begin{figure}[!htb]
	\center{\includegraphics[width=0.7\columnwidth]{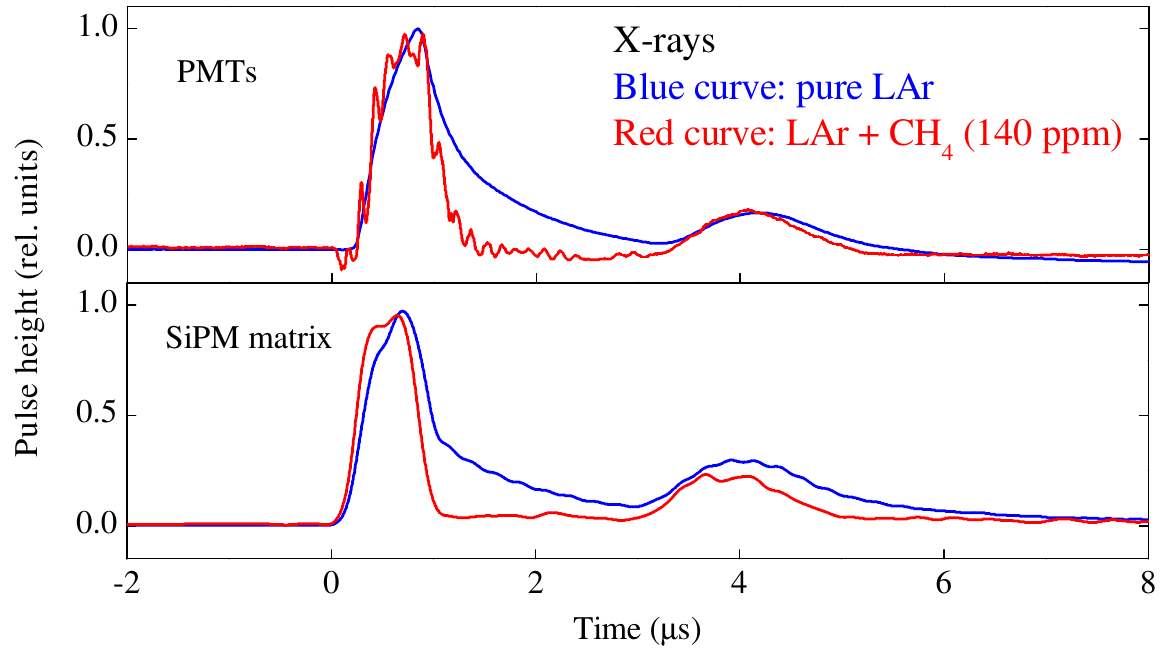}}
	\caption{Averaged pulse shapes of S1 signals, induced by pulsed X-rays, obtained in the two-phase mode in pure liquid Ar and liquid mixture Ar+CH$_4$ (140~ppm) at zero electric field, recorded by SiPM matrix and PMTs. All the pulse heights are normalized to the maximum.}
	\label{image:fig_pulse_shape_xray_noTPB}
\end{figure}

Fig.~\ref{image:fig_pulse_shape_xray_noTPB} shows the averaged pulse shapes of the S1 signals from PMTs and SiPM matrix, induced by pulsed X-rays, obtained in liquid Ar and liquid mixture Ar+CH$_4$ (140~ppm) at zero electric field. 
One can see that even a small amount of CH$_4$ dopant (140~ppm) resulted in disappearance of the slow component observed in pure liquid Ar.
With further increase of the CH$_4$ content, up to 10\%, the pulse shape did not change.
Also, the pulse shape did not depend on the electric field.

\begin{figure}[h!]
	\center{\includegraphics[width=0.85\columnwidth]{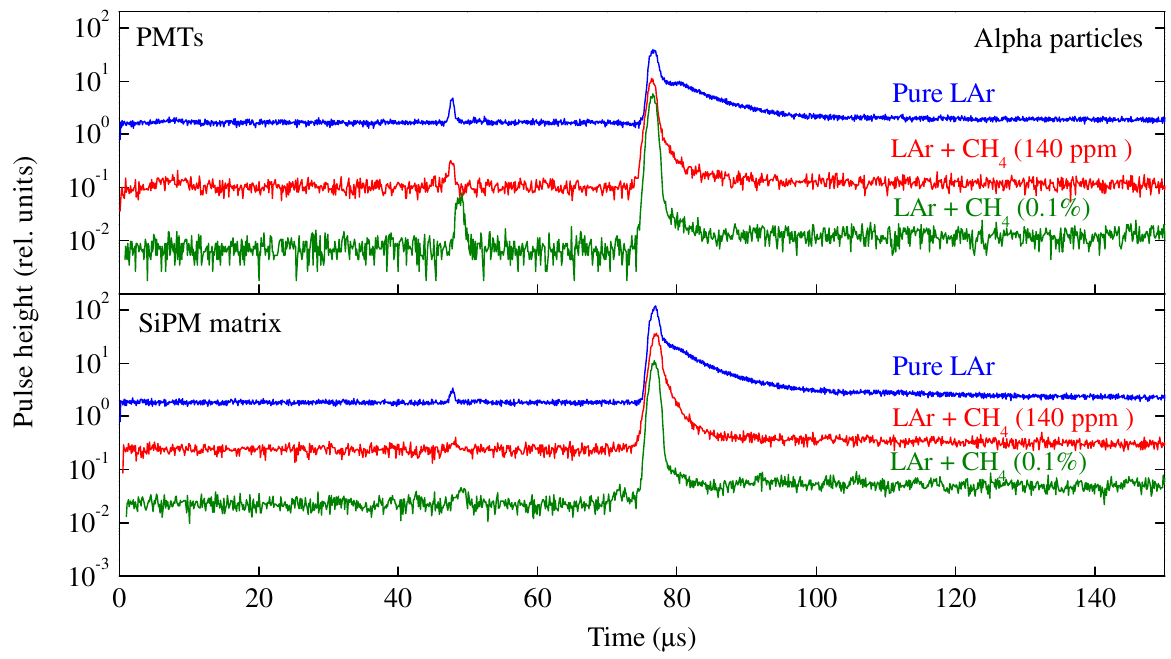}}
	\caption{Averaged pulse shapes of S1 and S2 signals, induced by alpha particles, obtained in the two-phase mode in pure liquid Ar and liquid mixtures Ar+CH$_4$ (140~ppm, 0.1\%) at maximum electric field (F$_d= 0.62$~kV/cm, F$_{EL} = 7.8$~kV/cm), recorded by SiPM matrix and PMTs. The pulse heights are proportional to the number of photoelectrons of S2 signal to highlight the difference in S2 pulse shapes and S2 yields. 
	}
	\label{image:fig_pulse_shape_S1_S2_alpha_noTPB}
\end{figure}

Fig.~\ref{image:fig_pulse_shape_S1_S2_alpha_noTPB} shows the averaged pulse shapes of both S1 and S2 signals, induced by alpha particles and obtained in the two-phase mode in pure liquid Ar and its mixtures with CH$_4$ (140~ppm, 0.1\% and 1\%) at the maximum electric field. The maximum electric field was needed to provide the efficient trigger using the S2 signal, as discussed in section 2. Accordingly, the S1 pulse width would be mostly defined by the jitter of the S2 pulse trigger, if the visible-light emission is really fast. We will see that this is indeed the case. 

It is seen that when increasing the CH$_4$ content the S2 signal became faster, its slow component quickly disappearing (see \cite{Bondar20S2} for explanation of the S2 slow component). In addition, its amplitude and drift time (time difference between S1 and S2) decreased. The estimated drift times in liquid Ar doped with methane are in good agreement with the data reported elsewhere (see fig.~\ref{fig_vd}).

\begin{figure}[h]
	\center{\includegraphics[width=0.7\columnwidth]{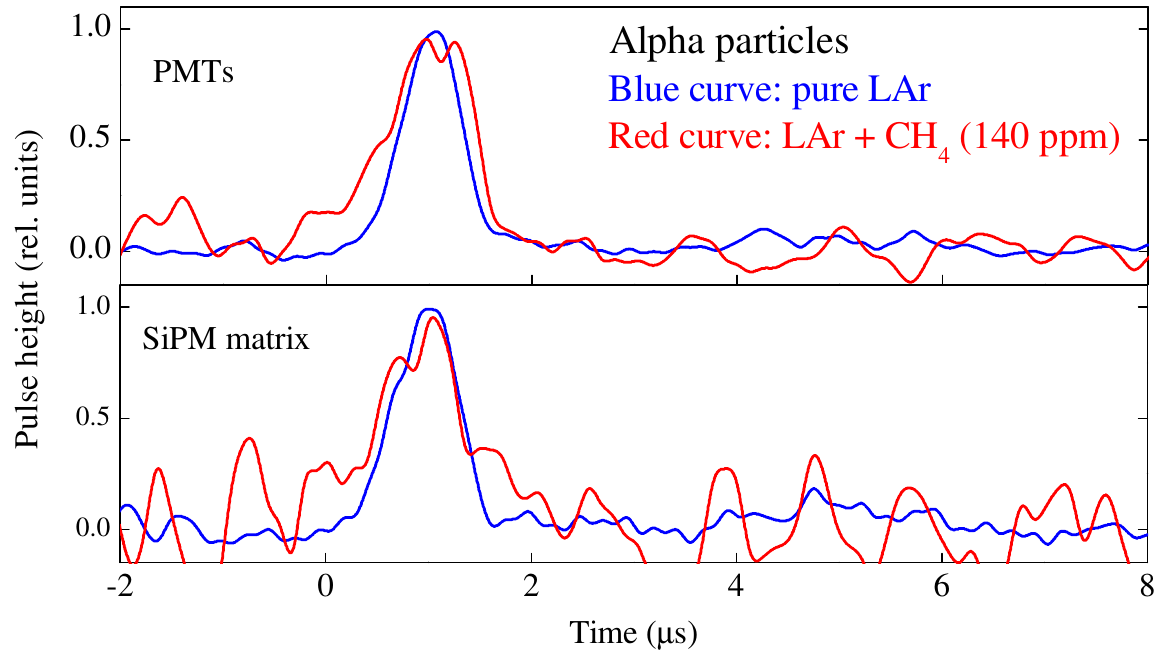}}
	\caption{Averaged pulse shapes of S1 signals, induced by alpha particles, obtained in the two-phase mode in pure liquid Ar and liquid mixture Ar+CH$_4$ (140~ppm) at the maximum electric field (F$_d= 0.62$~kV/cm, F$_{EL} = 7.8$~kV/cm), recorded by SiPM matrix and PMTs. All the pulse heights are normalized to the maximum.}
	\label{image:fig_pulse_shape_alpha_noTPB}
\end{figure}

Fig.~\ref{image:fig_pulse_shape_alpha_noTPB} shows the enlarged view of the S1 signals from Fig.~\ref{image:fig_pulse_shape_S1_S2_alpha_noTPB} in pure liquid Ar and its mixture with CH$_4$ (140~ppm).
One can see that in both cases the S1 pulse shapes look the same, containing only the fast component. This may indicate on the common origin of the fast component in visible-light emission in pure liquid Ar and in its mixtures with CH$_4$. \Xremove{We believe that this fast visible-light emission is due to the neutral bremsstrahlung effect.} \Xadd{Here the pulse shape of the fast component is defined mostly by the S2 signal jitter, which was much lager than the characteristic time of the NBrS emission in pure Ar and CH$_4$-doped Ar. That is why the S1 pulse shape was not sensitive to doping Ar with CH$_4$.}

\subsection{Photoelectron and light yields}

\begin{figure}[h!]
	\center{\includegraphics[width=0.8\columnwidth]{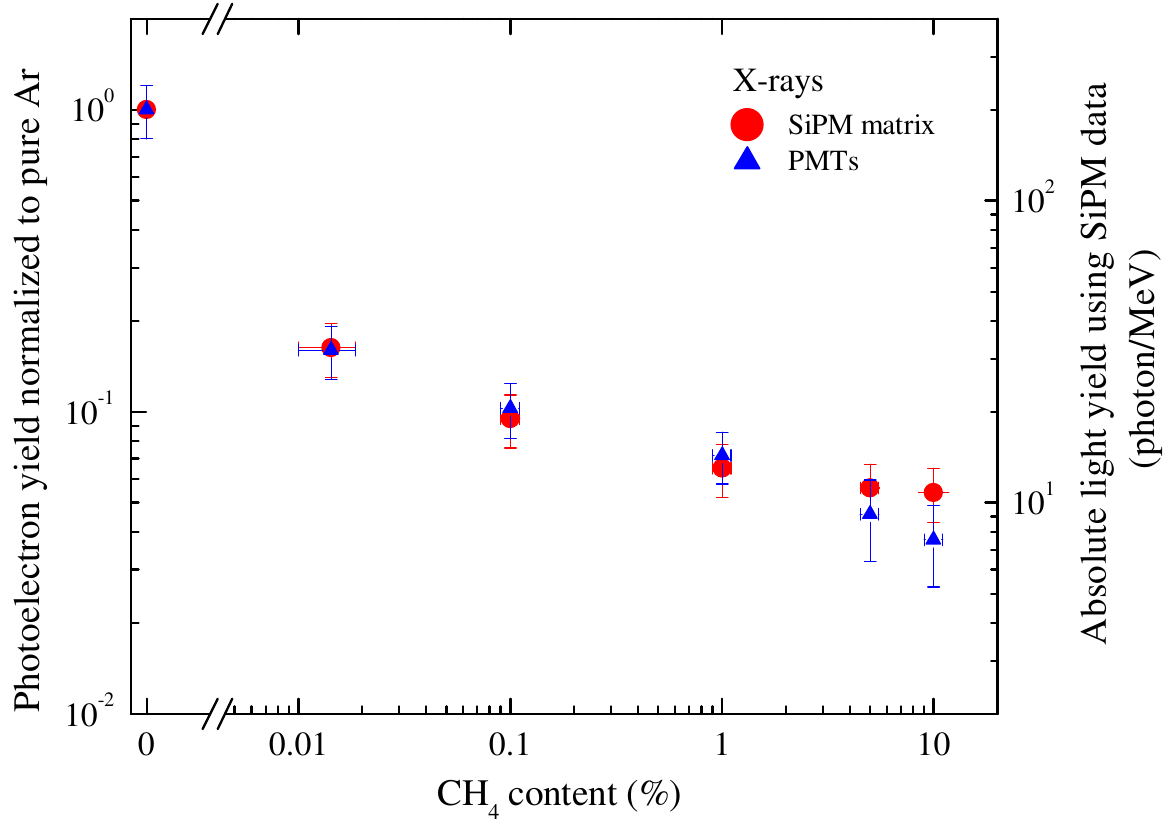}}
	\caption{Relative photoelectron yield of visible-light emission, induced by pulsed X-rays and measured in the two-phase mode in liquid Ar mixtures with CH$_4$, as a function of the CH$_4$ molar content at zero electric field, recorded by SiPM matrix and PMTs. Also shown is the absolute light yield obtained using SiPM-matrix data and averaged over two emission spectra from refs.~\cite{Auger16,Heindl11} (right scale, relevant to the SiPM-matrix data points).}
	\label{image:fig_rel_light_yield_xray_noTPB}
\end{figure}

\begin{figure}[h]
	\center{\includegraphics[width=0.75\columnwidth]{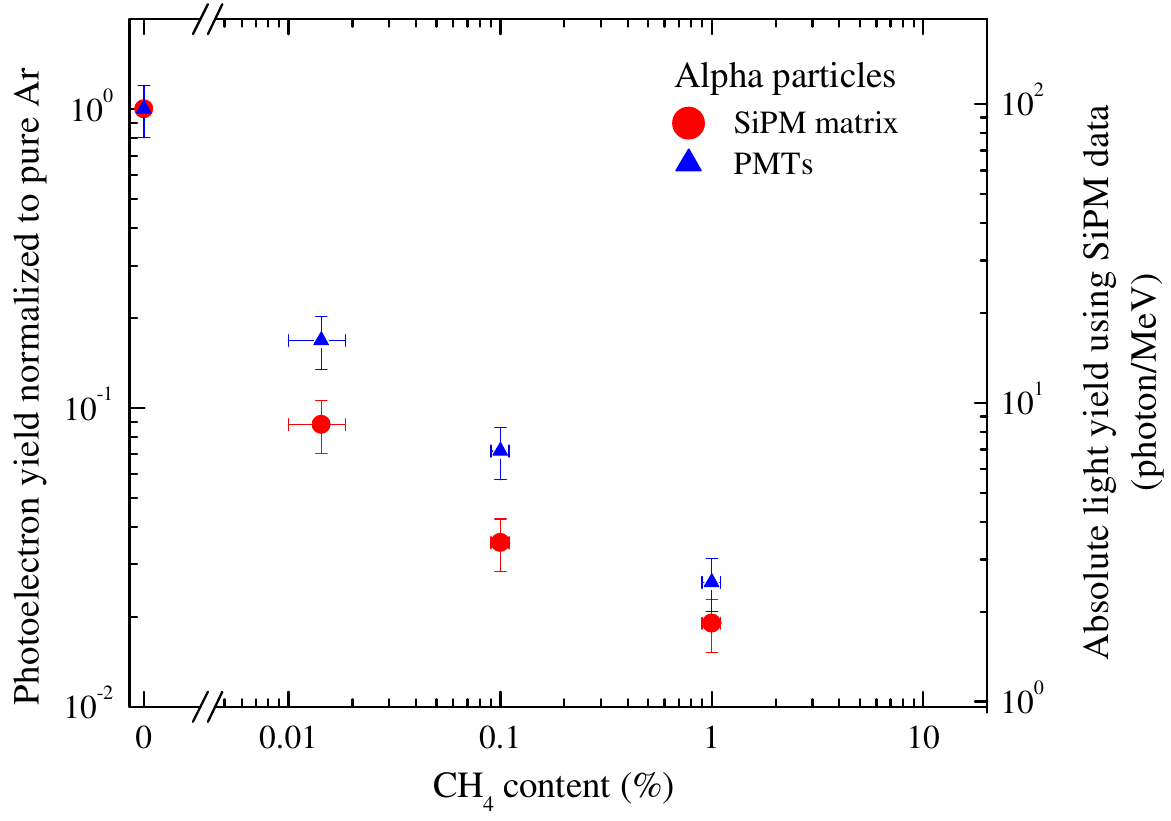}}
	\caption{Relative photoelectron yield of visible-light emission, induced by alpha particles and measured in the two-phase mode in liquid Ar mixtures with CH$_4$, as a function of the CH$_4$ molar content at the maximum electric field (F$_d= 0.62$~kV/cm, F$_{EL} = 7.8$~kV/cm), recorded by SiPM matrix and PMTs. Also shown is the absolute light yield obtained using SiPM-matrix data and averaged over two emission spectra from \cite{Auger16,Heindl11} (right scale, relevant to the SiPM-matrix data points).}
	\label{image:fig_rel_light_yield_alpha_noTPB}
\end{figure}

To calculate the photoelectron yield in argon-methane mixtures, the  photoelectron number was calculated in a 10~$\mu$s interval, similarly to that of pure Ar.
Figs.~\ref{image:fig_rel_light_yield_xray_noTPB} and \ref{image:fig_rel_light_yield_alpha_noTPB} show the relative photoelectron yield of  visible-light emission induced by pulsed X-rays and alpha particles in liquid Ar mixtures with CH$_4$ as a function of the CH$_4$ content. 
One can see that the data obtained with PMTs and SiPM matrix are in reasonable agreement, within a factor of 1.5 on average, despite the fact that PMTs and SiPMs are sensitive in different spectral ranges (see Fig.~\ref{fig_spectra}). This fact indicates that the spectrum of visible-light emission is rather flat and that its shape does not significantly change when CH$_4$ dopant is added. 

The absolute light yield is also shown in the figures~\ref{image:fig_rel_light_yield_xray_noTPB} and \ref{image:fig_rel_light_yield_alpha_noTPB} (right scale), obtained similarly to that of pure Ar, i.e. using SiPM-matrix data and averaged over the two emission spectra taken from \cite{Auger16,Heindl11}. Here the visible-light emission spectra of \cite{Auger16,Heindl11} for pure liquid Ar were used since they are the only ones available and since those for argon-methane mixtures are not available in the literature. Such a treatment is allowed due to the following points. Firstly, one could see in Section 3.4 that the photon yield was almost independent of the spectrum shape, in particular due to the fact that the spectra used are rather flat (see Fig.~\ref{fig_spectra}). Secondly, if the spectrum is indeed determined by the NBrS effect, its shape should not depend much on which species the electron is scattered on (Ar or CH$_4$), since numerous NBrS spectra of various origins presented in the literature look very flat and therefore very similar to each other~\cite{Buzulutskov18,Borisova21G,Borisova21L}. \Xadd{On the other hand, the presence of CH$_4$ dopant can affect the energy distribution of delta electrons and hence the properties of the NBrS emission produced by them, but our experimental procedures were not sensitive to such changes in the emission spectrum.}

\begin{figure}[h]
	\center{\includegraphics[width=0.7\columnwidth]{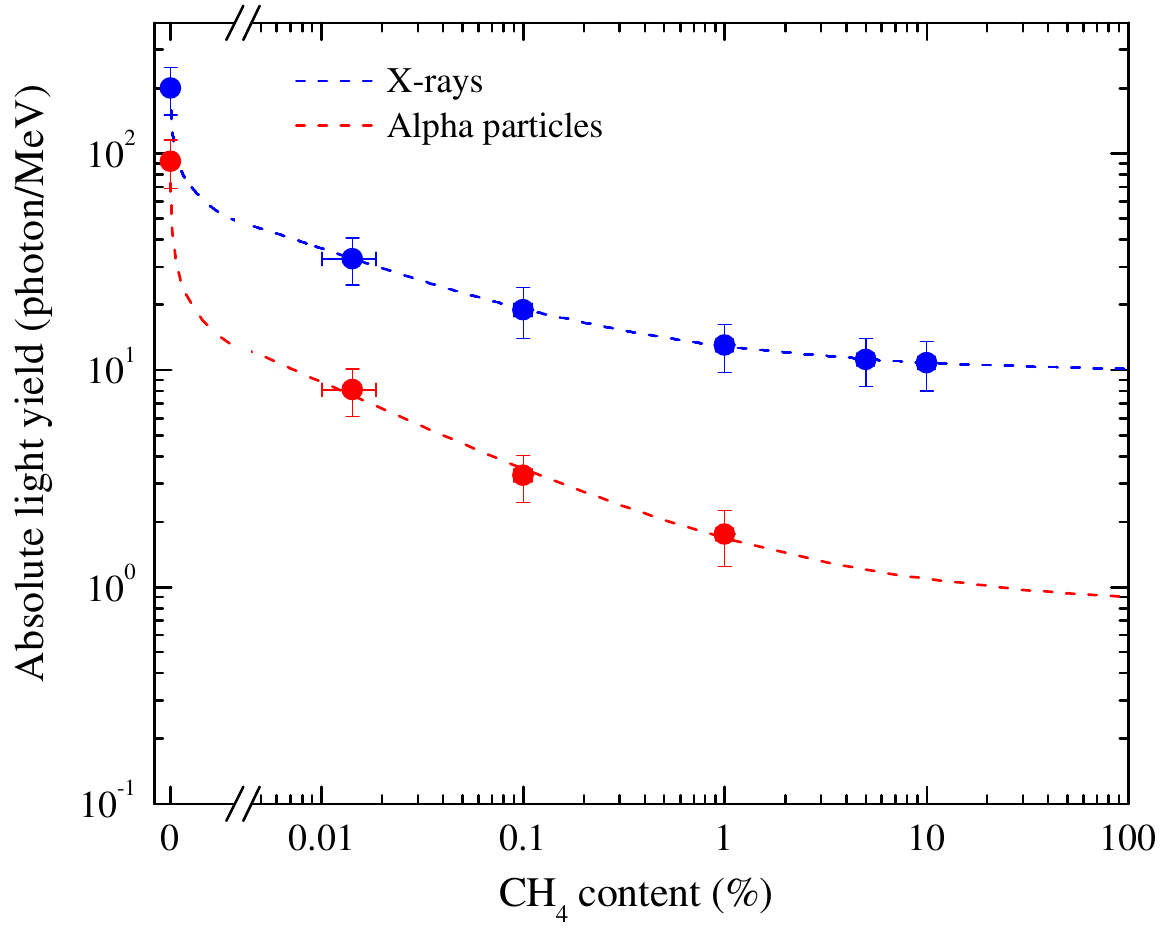}}
	\caption{Absolute light yield, induced by pulsed X-rays at zero electric field and by alpha particles at the maximum electric field (F$_d= 0.62$~kV/cm, F$_{EL} = 7.8$~kV/cm), in liquid Ar mixtures with CH$_4$ as a function of the CH$_4$ content. The results were obtained using SiPM-matrix data and averaged over two emission spectra from refs.~\cite{Auger16,Heindl11}. The curves are the sigmoidal function fit to the data points.}
	\label{image:fig_rel_light_yield_SiPM_alpha_vs_xray_noTPB}
\end{figure}

Fig.~\ref{image:fig_rel_light_yield_SiPM_alpha_vs_xray_noTPB} and table~\ref{table:avr_abs_yields} summarize the results on the absolute photon yield of visible-light emission in pure liquid Ar and its mixtures with CH$_4$. Looking at these results one may conclude the following.

For X-rays, the absolute photon yield of visible-light emission in pure liquid Ar is about 200 photon/MeV. In liquid argon doped with methane, the photon yield dropped down significantly, by about an order of magnitude at a CH$_4$ content varying from 0.01 to 1\%, and then almost did not change when further increasing the content up to 10\%, reaching a plateau at about 10 photon/MeV.


For alpha particles, the photon yield is reduced compared to that of X-rays, the reduction factor changing from 2 in pure liquid Ar (90 photon/MeV) to 7 at 1\% CH$_4$ content. Nevertheless, the CH$_4$ content  dependence looks very similar to that of X-rays: there is an obvious trend towards a plateau, but due to the lack of data above 1\% it is not completed (at higher CH$_4$ contents the S2 signal was too weak to provide the trigger). Extrapolation of the data gives a plateau value of about 1 photon/MeV for the photon yield.

Such a characteristic dependence on the CH$_4$ content can be qualitatively explained in the framework of NBrS mechanism. Indeed, the NBrS intensity is proportional to the energy of primary ionization electrons (see formula 4 in \cite{Buzulutskov18}). When adding CH$_4$ to liquid Ar, the electron energy decreases with the CH$_4$ content due to enhancement of inelastic collisions with  CH$_4$ molecules, until they fully dominate at higher contents, thus producing a plateau in CH$_4$ content dependence.  

Finally, with a CH$_4$ content of 1\% (the minimum required for a compact neutron veto detector), the absolute photon yield of visible-light emission for X-rays is about 13~photon/MeV. This value seems to be too small for the effective detection of gamma-rays accompanying neutron capture on argon and hydrogen. Accordingly, liquid Ar doped with CH$_4$ can hardly be used as a scintillating medium for neutron veto detectors. On the other hand, it might be considered to be used in scintillation hadron calorimetry, since methane allows to compensate the calorimeter and thus increase its energy resolution.

\begin{table}[h]
	\caption{
		Absolute light yield in pure liquid Ar and its mixtures with CH$_4$ at 400-1000~nm induced by pulsed X-rays and alpha particles at zero and maximum electric field (F$_d= 0.62$~kV/cm, F$_{EL} = 7.8$~kV/cm) respectively.}\label{table:avr_abs_yields}
	\begin{center}
		\begin{tabular}{|ccc|}
			\hline
			\multirow{2}{*}{Liquid mixture} & \multicolumn{2}{c|}{Light yield  (photon/MeV)} \\
			& \makecell{Pulsed X-rays \\ (zero field) } & \makecell{Alpha particles \\ (F$_d=0.62$~kV/cm)} \\
			\hline
			LAr~(100\%) & 200$\pm$50 & 92$\pm$23 \\
			LAr + CH$_4$~(140 ppm ) & 32$\pm$8 & 8.2$\pm$2.0 \\
			LAr + CH$_4$~(0.1\% ) & 19$\pm$5 & 3.3$\pm$0.8 \\
			LAr + CH$_4$~(1\%) & 13.0$\pm$3.3 & 1.8$\pm$0.5 \\
			LAr + CH$_4$~(5\%) & 11.3$\pm$2.8 & -  \\
			LAr + CH$_4$~(10\%) & 11.1$\pm$2.8 & -  \\
			\hline
		\end{tabular}
	\end{center}
\end{table}

\section{Conclusions}

In this work the properties of visible-light emission in pure liquid argon and its mixtures with methane have been systematically studied, using cryogenic PMTs and a SiPM matrix. The light yield in pure liquid argon was measured to be about 200 and 90 photon/MeV for X-rays and alpha particles respectively, in the wavelength range of 400-1000 nm. 

In liquid argon doped with methane the light yield dropped down significantly, by about an order of magnitude at a methane molar content varying from 0.01 to 1\%, and then almost did not change when further increasing the methane content up to 10\%, reaching a plateau at about 10 photon/MeV for X-rays.  

For alpha particles, the light yield is reduced compared to that of X-rays, the reduction factor changing from 2 in pure liquid argon to 7 at 1\% methane molar content.

Due to such a small light yield, liquid argon doped with methane can hardly be used as working medium for neutron veto detectors. \Xremove{On the other hand, it might be used in scintillation hadron calorimetry, since methane allows to compensate the calorimeter and thus increase its energy resolution.}

\Xadd{Based on the observed properties of visible-light emission in both pure and methane-doped liquid argon, we propose that it is due to neutral bremsstrahlung of primary ionization electrons.}

\appendix
\section{Doping liquid argon with methane: drift velocity and charge yield}

In this Appendix a compilation of data on drift velocity and charge yield in liquid Ar, liquid CH$_4$ and their mixtures is presented.  

\begin{figure}[!htb]
	\center{\includegraphics[width=0.7\columnwidth]{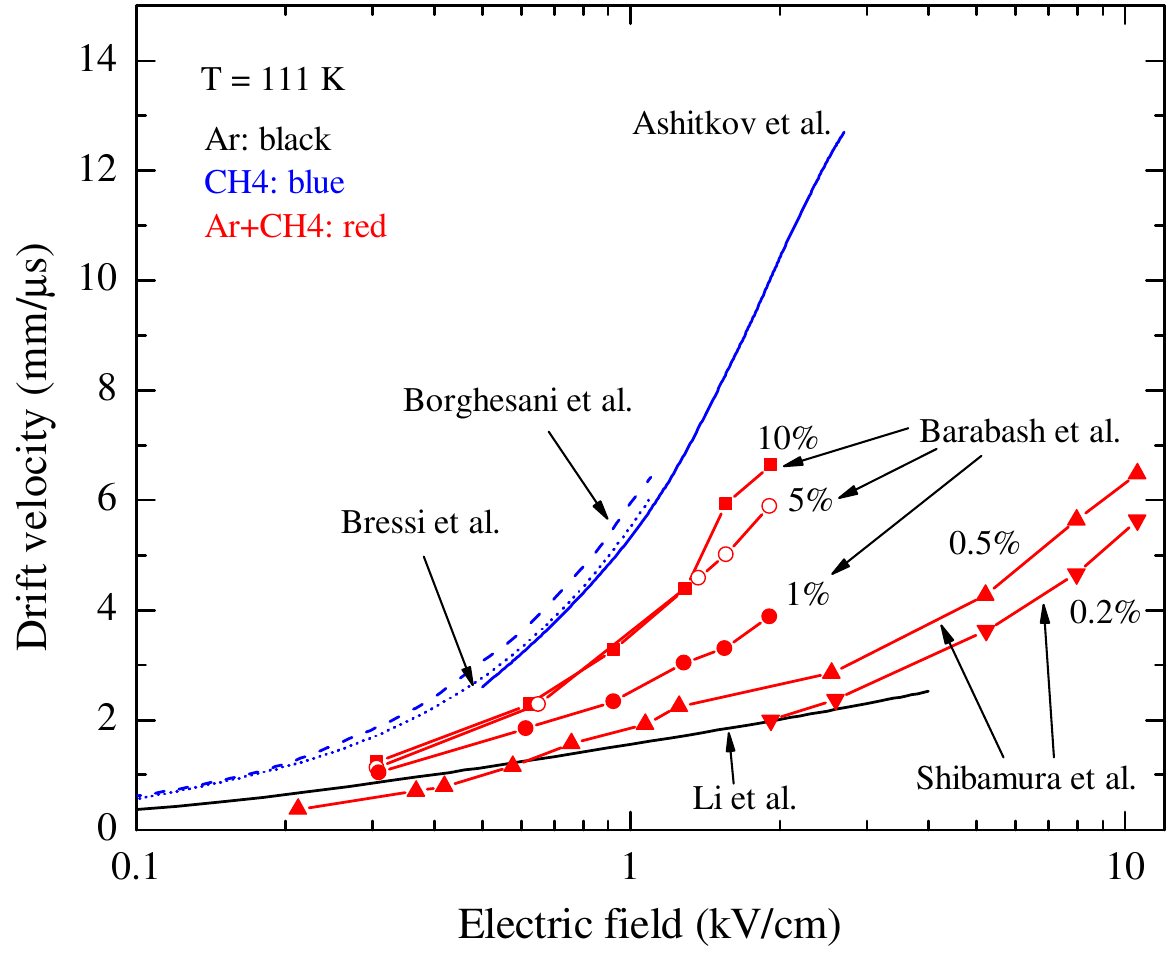}}
	\caption{Electron drift velocity as a function of the electric field in liquid Ar, liquid CH$_4$ and their mixtures taken from Ashitkov et al.~\cite{Ashitkov01}, Barabash et al.~\cite{Barabash81}, Borghesani et al.~\cite{Borghesani91}, Bressi et al.~\cite{Bressi91}, Li et al.~\cite{Li16}, Shibamura et al.~\cite{Shibamura75}. The drift velocity values are reduced to the same temperature, of 111~K, using $T^{-3/2}$ dependence \cite{Cohen67,Engels77,Li16}.}
	\label{fig_vd}
\end{figure}

Fig.~\ref{fig_vd} shows the electron drift velocity as a function of the electric field in these media.
Since the drift velocity depends on the liquid temperature as $T^{-3/2}$ over a wide temperature range~\cite{Cohen67,Engels77,Li16}, all data are reduced to the same temperature, of 111~K, for comparison under the same conditions. The temperature of 111~K was chosen to minimize the interpolation error and because it corresponds to the boiling point of methane at atmospheric pressure. One can see that the electron drift velocity substantially increases with the CH$_4$ content.

\begin{figure}[!htb]
	\center{\includegraphics[width=0.7\columnwidth]{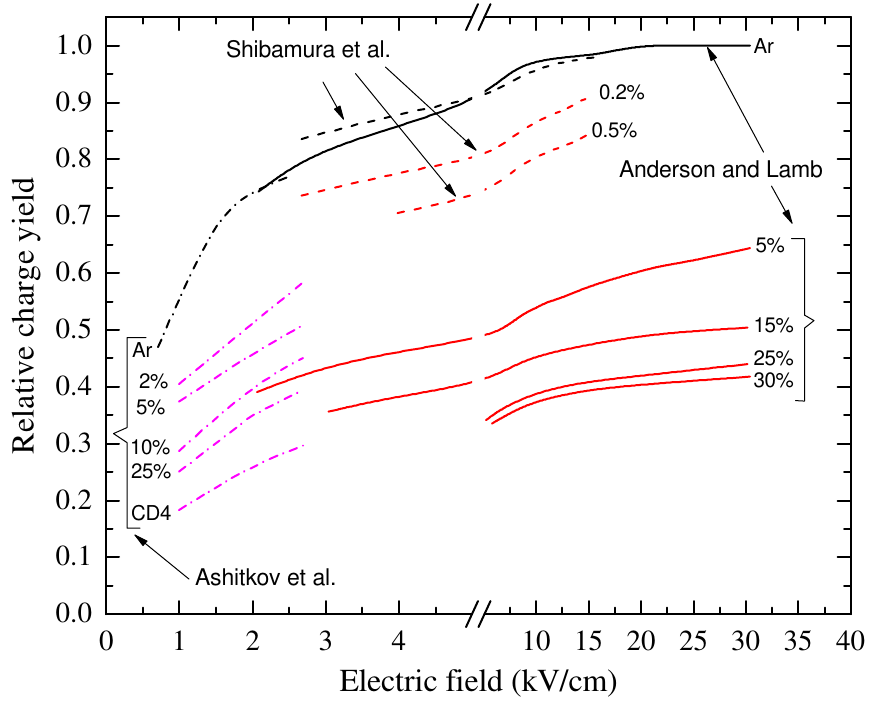}}
	\caption{Relative charge yield from tracks of relativistic electrons as a function of the electric field in liquid Ar and its mixtures with CH$_4$/CD$_4$ taken from Anderson and Lamb~\cite{Anderson88} (maximum electron energy of 3.5~MeV), Ashitkov et al.~\cite{Ashitkov01} (1.06~MeV Compton electrons), Shibamura et al.~\cite{Shibamura75} (0.976~MeV electrons),  normalized to the maximum collected charge in liquid Ar at the infinite electric field.}
	\label{fig_charge_yield_beta}
\end{figure}

\begin{figure}[!htb]
	\center{\includegraphics[width=0.7\columnwidth]{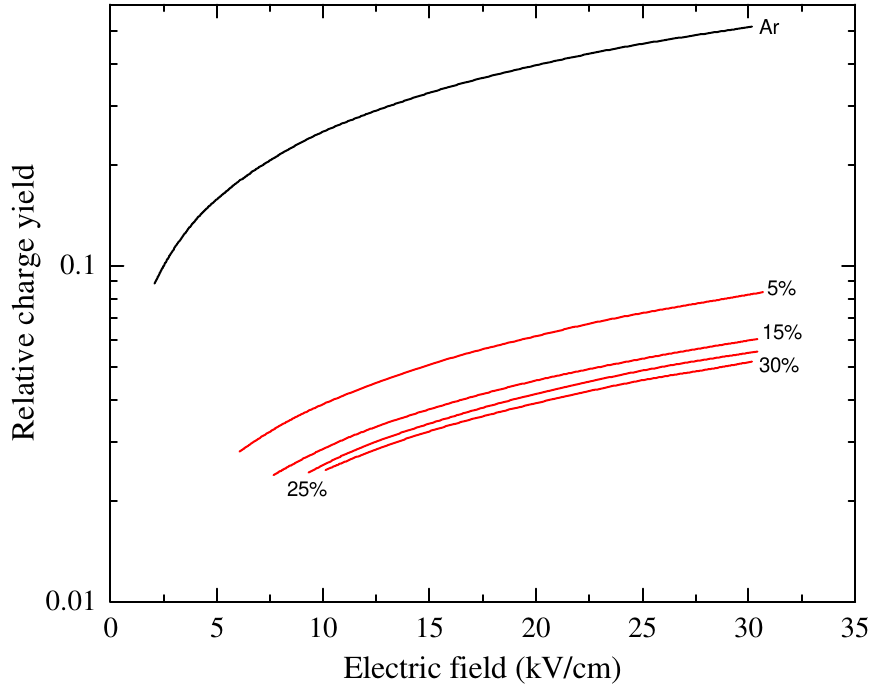}}
	\caption{Relative charge yield from tracks of 5.5~MeV alpha particles as a function of the electric field in liquid Ar and its mixtures with CH$_4$ at CH$_4$ content of 5\%, 15\%, 25\% and 30\%. The data are taken from Anderson and Lamb~\cite{Anderson88}, normalized to the maximum charge collected from tracks of relativistic electrons in liquid Ar at the infinite electric field.}
	\label{fig_charge_yield_alpha}
\end{figure}

Figs.~\ref{fig_charge_yield_beta} and \ref{fig_charge_yield_alpha} show the dependence of the relative charge yield, from tracks of relativistic electrons and alpha particles respectively, on the electric field in liquid Ar, liquid methane/deuteromethane and their mixtures. The relative charge yield is defined as the ratio of the charge collected at a given electric field to the maximum charge collected from tracks of relativistic electrons at the infinite electric field. 
Here the equation~\eqref{eq_charge_vs_field_full} with $T_e=100\%$ and $C=0$ was used to fit the charge yield as a function of the electric field. It is seen that the charge yield substantially decreases with the methane content.


\clearpage
\acknowledgments
This work was supported in part by Russian Science Foundation (project no. 20-12-00008).
It was done within the R\&D program of the DarkSide-20k experiment.

\bibliographystyle{JHEP}
\bibliography{mybibliography}

\end{document}